\documentclass[conference]{IEEEtran}

\IEEEoverridecommandlockouts   

\usepackage[mode=buildmissing]{standalone} % mode=image|tex, mode=build, buildmissing
\usepackage{amsmath}
\usepackage{bm}
\usepackage[nolist]{acronym}
\usepackage{amssymb}
\usepackage{comment}
\usepackage{graphicx}
\usepackage{color,colortbl}
\usepackage{xcolor}
\usepackage{cleveref}
\usepackage{verbatim}
\usepackage{enumerate}
\usepackage[shortlabels]{enumitem}
\usepackage{cuted, nccmath}
\usepackage{dblfloatfix}
\usepackage{float} 
\usepackage{lipsum}
\usepackage{balance}
\usepackage{cite}
\usepackage{url}
\usepackage{tabularx}
\usepackage{boldline}
\usepackage{balance}    
\usepackage{mathtools} % for DeclarePairedDelimiter
\usepackage{stmaryrd} % for mapsfrom
\usepackage{graphicx}
\usepackage{pgfplots}
\usepackage{booktabs}
\usepackage[font={footnotesize}]{caption}
\usepackage{subfig} 
\usepackage{tikz}
\usepackage{xparse} %To define commands with If statements
\usepackage{algorithm}
\usepackage{algpseudocode}

\usetikzlibrary{arrows.meta}
\tikzset{every picture/.style={line width=0.6pt}}
\usetikzlibrary{arrows.meta,
                calc, 
                backgrounds,
                chains,
                fit,
                quotes,
                shapes.geometric,
                positioning,
                intersections,
                spath3,
                spy}

\algdef{SE}[SUBALG]{Indent}{EndIndent}{}{\algorithmicend\ }%
\algtext*{Indent}
\algtext*{EndIndent}

\definecolor{myred}{rgb}{0,0,0}

\newcommand{\Aperturbed}{\bm{A}_{\rm{pert}}}
\newcommand{\admap}{\bm{\mathcal{L}}(\Yrtilde)}
\newcommand{\admapindex}[2]{[\bm{\mathcal{L}}(\Yrtilde)]_{#1,#2}}
\newcommand{\admapreduced}{ \overline{\bm{\mathcal{L}}} } 
\newcommand{\admapreducedsoft}{ \overline{\bm{\mathcal{L}}}_{\rm{soft}} } 
\newcommand{\aperturbed}{\bm{a}_{\rm{pert}}}
\newcommand{\arx}{\bm{a}_{\rm{rx}}}
\newcommand{\atx}{\bm{a}_{\rm{tx}}}

\newcommand{\bmAtx}{\bm{A}_{\rm{tx}}}
\newcommand{\bbold}{\bm{b}}

\newcommand{\bmbeta}{\bm{\beta}}
\newcommand{\bmd}{\bm{d}}
\newcommand{\bmF}{\bm{F}}
\newcommand{\bmf}{\bm{f}}
\newcommand{\bmfbs}{\bm{f}_{\rm{bs}}}
\newcommand{\bmI}{\bm{I}}
\newcommand{\bmkappa}{\bm{\kappa}}
\newcommand{\bmm}{\bm{m}}
\newcommand{\bmmhat}{\hat{\bm{m}}}
\newcommand{\bmN}{\bm{N}}
\newcommand{\bmn}{\bm{n}}
\newcommand{\bmPhi}{\bm{\Phi}}
\newcommand{\bmp}{\bm{p}}
\newcommand{\bmphat}{\hat{\bm{p}}}
\newcommand{\bmrho}{\bm{\rho}}
\newcommand{\bmvarthetainterval}{\bm{\vartheta}_{\rm{interval}}}
\newcommand{\bmW}{\bm{W}}
\newcommand{\bmx}{\bm{x}}
\newcommand{\boldone}{\boldsymbol{1}}
\newcommand{\hermit}{\mathsf{H}}

\newcommand{\cnormal}{\mathcal{CN}}
\DeclareDocumentCommand{\complexset}{o o}{%
    \mathbb{C}\IfValueT{#1}{\IfValueTF{#2}{^{#1\times#2}}{^{#1}}}
}
\newcommand{\conj}{ {\ast} }

\newcommand{\Deltaf}{\Delta_f}
\newcommand{\degree}{^{\circ}}
\newcommand{\deltatheta}{\delta_{\theta}}
\newcommand{\deltaR}{\delta_{R}}

\newcommand{\E}{\mathbb{E}}

\newcommand{\ihat}{\hat{i}}

\newcommand{\Jsuper}{\mathcal{J}_{\text{SL}}}
\newcommand{\Junsuper}{\mathcal{J}_{\text{UL}}}
\newcommand{\jhat}{\hat{j}}

\newcommand{\hypz}{ \mathcal{H}_0 }
\newcommand{\hypone}{ \mathcal{H}_1 }
\newcommand{\hdet}{ \underset{\mathcal{H}_0}{\overset{\mathcal{H}_1}{\gtrless}} }

\newcommand{\Mset}{\mathcal{M}}

\newcommand{\Ntheta}{N_{\theta}}
\newcommand{\normal}{\mathcal{N}}
\newcommand{\nubar}{\widebar{\nu}}

\newcommand{\Pfa}{P_{\rm{fa}}}
\newcommand{\Pmd}{P_{\rm{md}}}
\renewcommand{\Pr}{p}

\newcommand{\Rhat}{\hat{R}}
\newcommand{\Rmax}{R_{\max}}
\newcommand{\Rmin}{R_{\min}}
\DeclareDocumentCommand{\realset}{o o}{%
    \mathbb{R}\IfValueT{#1}{\IfValueTF{#2}{^{#1\times#2}}{^{#1}}}
}
\newcommand{\sigmalambda}{\sigma_{\lambda}}
\newcommand{\SNRr}{\mathrm{SNR}_r} %WIth just \rm it changes all the text after this command
\newcommand{\SNRc}{\mathrm{SNR}_c}

\newcommand{\that}{\hat{t}}
\newcommand{\thetagrid}{\{\theta_i\}_{i=1}^{\Ntheta}}
\newcommand{\thetahat}{\widehat{\theta}}
\newcommand{\taugrid}{\{\tau_i\}_{i=1}^{N_{\tau}}}
\newcommand{\tauhat}{\widehat{\tau}}
\newcommand{\thetamax}{\theta_{\max}}
\newcommand{\thetamean}{\theta_{\rm{mean}}}
\newcommand{\thetamin}{\theta_{\min}}
\newcommand{\thetaspan}{\theta_{\rm{span}}}

\newcommand{\taumax}{\tau_{\max}}
\newcommand{\taumin}{\tau_{\min}}

\newcommand{\U}{\mathcal{U}}

\newcommand{\varphimax}{\varphi_{\max}}
\newcommand{\varphimin}{\varphi_{\min}}
\newcommand{\varthetamax}{\vartheta_{\max}}
\newcommand{\varthetamin}{\vartheta_{\min}}
\newcommand{\vecc}[1]{ {\rm{vec}}(#1)  }

\newcommand{\yc}{\bm{y}_c}
\newcommand{\Yr}{\bm{Y}_r}
\newcommand{\Yrtilde}{\widetilde{\bm{Y}}_r}

\DeclarePairedDelimiter\absbigs{\big\lvert}{\big\rvert}%

\DeclarePairedDelimiter\norm{\lVert}{\rVert}%

\makeatletter
\newcommand*\rel@kern[1]{\kern#1\dimexpr\macc@kerna}
\newcommand*\widebar[1]{%
  \begingroup
  \def\mathaccent##1##2{%
    \rel@kern{0.8}%
    \overline{\rel@kern{-0.8}\macc@nucleus\rel@kern{0.2}}%
    \rel@kern{-0.2}%
  }%
  \macc@depth\@ne
  \let\math@bgroup\@empty \let\math@egroup\macc@set@skewchar
  \mathsurround\z@ \frozen@everymath{\mathgroup\macc@group\relax}%
  \macc@set@skewchar\relax
  \let\mathaccentV\macc@nested@a
  \macc@nested@a\relax111{#1}%
  \endgroup
}
\makeatother

\begin{acronym}[ACRONYM]
\acro{6G}{6th generation wireless systems}

\acro{AE}{autoencoder}
\acro{AoA}{angle-of-arrival}
\acro{AoD}{angle-of-departure}
\acro{AWGN}{additive white gaussian noise}

\acro{BCE}{binary cross-entropy}

\acro{CCE}{categorial cross-entropy}
\acro{CNN}{convolutional neural network}
\acro{CSI}{channel state information}

\acro{DFT}{discrete Fourier transform}
\acro{DL}{deep learning}
\acro{DNN}{deep neural network}
\acro{DOA}{direction of arrival}

\acro{GNSS}{global navigation satellite system}
\acro{GOSPA}{generalized optimal sub-pattern assignment}

\acro{IoT}{internet of things}
\acro{IRS}{intelligent reflecting surface}
\acro{ISTA}{iterative shrinkage-thresholding algorithm}
\acro{ISAC}{integrated sensing and communication}

\acro{JRC}{joint radar and communication}

\acro{LS}{least squares}

\acro{MAP}{maximum a posteriori} 
\acro{MB-ML}{model-based machine learning}
\acro{MIMO}{multiple-input multiple-output}
\acro{MISO}{multiple-input single-output}
\acro{ML}{machine learning} 
\acro{MLE}{maximum likelihood estimation} 
\acro{MLP}{multilayer perceptron}
\acro{MSE}{mean squared error}

\acro{NLL}{negative log-likelihood}
\acro{NN}{neural network}
\acro{NNBL}{neural-network-based learning}

\acro{OFDM}{orthogonal frequency-division multiplexing}
\acro{OMP}{orthogonal matching pursuit}

\acro{PDF}{probability density function}

\acro{QPSK}{quadrature phase shift keying}

\acro{ReLU}{rectified linear unit}
\acro{RIS}{reconfigurable intelligent surface}
\acro{RMSE}{root mean squared error}
\acro{ROC}{receiver operating characteristic}
\acro{RSS}{received signal strengh}

\acro{SER}{symbol error rate}
\acro{SL}{supervised learning}
\acro{SNR}{signal-to-noise ratio}
\acro{SSL}{semi-supervised learning}

\acro{UE}{user equipment}
\acro{UL}{unsupervised learning}
\acro{ULA}{uniform linear array}

\acro{V2I}{vehicle-to-infrastructure}
\end{acronym}

\begin{document}
\bstctlcite{IEEEexample:BSTcontrol}

% the \LARGE sets the title at 17 pt, which is close enough to 18-point.
%+++++++++++++++++++++++++++++++++++++++++++
\title{Semi-Supervised End-to-End Learning for Integrated Sensing and Communications \\
\thanks{This work was supported, in part, by a grant from the Chalmers AI Research Center Consortium (CHAIR), by the National Academic Infrastructure for Supercomputing in Sweden (NAISS), the Swedish Foundation for Strategic Research (SSF) (grant FUS21-0004, SAICOM), Hexa-X-II, part of the European Union’s Horizon Europe research and innovation programme under Grant Agreement No 101095759, and Swedish Research Council (VR grant 2022-03007). The work of C.~Häger was also supported by the Swedish Research Council under grant no. 2020-04718.}
}
\author{Jos\'{e} Miguel Mateos-Ramos\IEEEauthorrefmark{1}, Baptiste Chatelier\IEEEauthorrefmark{2}, Christian H\"{a}ger\IEEEauthorrefmark{1}, \\Musa Furkan Keskin\IEEEauthorrefmark{1}, Luc Le Magoarou\IEEEauthorrefmark{2}, Henk Wymeersch\IEEEauthorrefmark{1}\\
\IEEEauthorrefmark{1}Department of Electrical Engineering, Chalmers University of Technology, Sweden \\
\IEEEauthorrefmark{2}Univ Rennes, INSA Rennes, CNRS, IETR-UMR 6164, Rennes, France}

\maketitle

\IEEEpeerreviewmaketitle

\begin{abstract}
Integrated sensing and communications (ISAC) is envisioned as one of the key enablers of next-generation wireless systems, offering improved hardware, spectral, and energy efficiencies. 
In this paper, we consider an ISAC transceiver with an impaired uniform linear array that performs single-target detection and position estimation, and multiple-input single-output communications. A differentiable model-based learning approach is considered, which optimizes both the transmitter and the sensing receiver in an end-to-end manner. An unsupervised loss function that enables impairment compensation without the need for labeled data is proposed. Semi-supervised learning strategies are also proposed, which use a combination of small amounts of labeled data and unlabeled data. Our results show that semi-supervised learning can achieve similar performance to supervised learning with 98.8\% less required labeled data.
\end{abstract}


\section{Introduction} \label{sec_introduction}
\Ac{ISAC} refers to the combination of sensing and communication resources to improve the efficiency or performance of a system, or to endow a system with new functionalities. In \ac{ISAC} systems, communication and sensing tasks share the same spectrum, hardware, or signal processing algorithms, which enhances hardware, spectral or energy efficiency \cite{cui2023integrated}. Emerging technologies such as digital twins, activity recognition, and extended reality are largely driven by \ac{ISAC} \cite{wymeersch2021integration,liu2022integrated,lu2023integrated}, placing it as one of the key enablers of next-generation wireless systems.

ISAC designs have largely relied on model-based signal processing algorithms, which offer performance guarantees, explainability, and predictable computational complexity\textcolor{myred}{\cite{liyanaarachchi2021joint, dokhanchi2021adaptive, OFDM_DFRC_TSP_2021, chen2021joint, johnston2022mimo}}. However, under unexpected hardware impairments such as antenna distortions, sampling jitter, or miscalibration errors \cite{tan2021integrated, chowdhury20206g, jiang2021road}, model-based algorithms can severely degrade in performance.
With the advent of \acp{DNN}, hardware impairments and mismatched models have been recently tackled by \ac{DL}, which in the context of ISAC can be categorized in: (i) single-component \ac{DL}, where a \ac{DNN} is optimized to perform transmit or receive operations individually \cite{liu2022learning, wu2023sensing, liu2023deep}, and (ii) end-to-end \ac{DL}, where both tasks are implemented as \acp{DNN}, forming an \ac{AE} architecture \cite{OShea2017}, and jointly optimized \cite{mateos2022end, muth2023autoencoder}. 
\acp{DNN} are essentially \emph{black boxes}, limiting the interpretability of the learned function. Moreover, they usually require large amounts of labeled training data.
\begin{figure}[tb]
    \centering
    \includegraphics[width=0.47\textwidth]{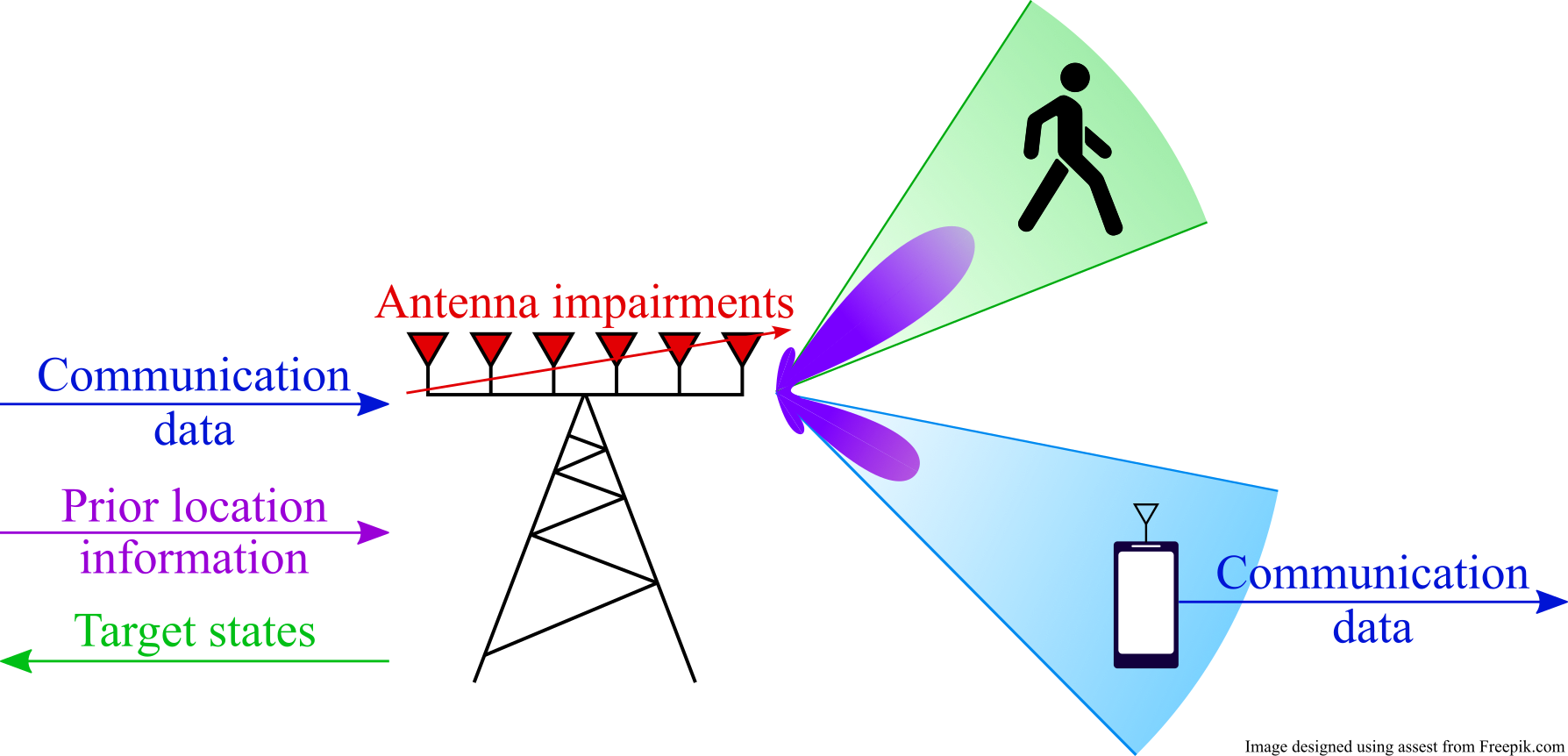}
    \caption{Considered ISAC scenario. An impaired ISAC transceiver senses a single target in the environment while communicating with a device in another location. The transceiver is assumed to have some prior knowledge about the approximate location of the target and the communication receiver.}
    \label{fig_scenario}
\end{figure}

A way to overcome the limited interpretability of DNNs is to follow the \ac{MB-ML} paradigm~\cite{Shlezinger2023}, whereby existing models from signal processing are used to initialize, structure, and train learning methods. 
Furthermore, since MB-ML relies on problem-specific architectures, it generally requires less labeled data than model-agnostic DNNs. Recently, MB-ML has been applied in several communication~\cite{Xiuhong21,Hengtao18, chatelier2023modelbased}, sensing \cite{xiao2020deepfpc, wu2022doa}, and ISAC \cite{mateos2023model} scenarios. Nevertheless, previous MB-ML approaches \cite{Xiuhong21, Hengtao18, chatelier2023modelbased, xiao2020deepfpc, wu2022doa, mateos2023model} rely on labeled data to train the MB-ML models, which can be hard or time-consuming to acquire in sensing environments, especially for automotive sensing.
Further reducing the need of labeled data is possible through \ac{SSL} or \ac{UL}. UL performs learning just based on the observed (unlabeled) data, while SSL combines a limited amount of labeled data with unlabeled data \cite{chapelle2006ssl}. 
Merging both MB-ML and UL approaches has been applied in the context of channel estimation for communications~\cite{yassine2022,Chatelier2022}. However, the use of unlabeled data in MB-ML for ISAC remains unexplored. This topic deserves particular attention, since labeled data in sensing environments involves acquiring the ground-truth position of all targets in the scene through external sensors (e.g., cameras or global navigation satellite system), which might become a challenging task. 

In this paper, we propose a novel approach that enables \ac{UL} and \ac{SSL} in an \ac{ISAC} scenario. We consider a monostatic \ac{MIMO} radar to perform single-target estimation and a \ac{MISO} communication link, as shown in Fig.~\ref{fig_scenario}. 
This is a special case of the multi-target estimation scenario recently studied in \cite{mateos2023model}. Compared to \cite{mateos2023model} and other end-to-end learning works on ISAC \cite{mateos2022end, muth2023autoencoder}, the main novelty of this work is the use of unlabeled data to reduce the acquisition of labeled data, while maintaining performance similar to the fully-supervised case. We propose a corresponding unsupervised loss function and perform a comprehensive study comparing: (i) a model-aware baseline that involves no learning, (ii) \ac{SL}, (iii) \ac{UL}, and (iv) \ac{SSL} with different degrees of labeled data. 

\section{System Model}  \label{sec_model}
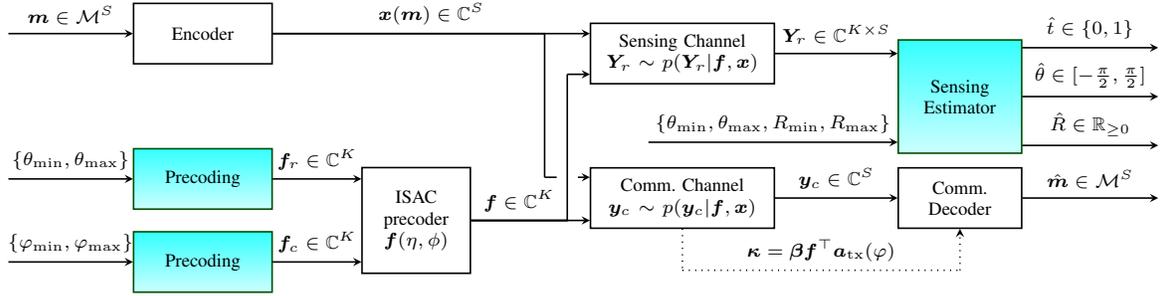
\begin{figure*}[t]
    \centering
    \begin{tikzpicture}[font=\scriptsize, >=stealth, blkCH/.style={draw,minimum height=0.8cm,text width=2.2cm, text centered},  blk/.style={draw,minimum height=0.8cm,text width=1.4cm, text centered}, blkNN/.style={shading=axis, left color=cyan!80, right color=cyan!20, shading angle=0, draw=green!30!black,minimum height=0.8cm,text width=1.4cm, text centered},  x=0.6cm,y=0.55cm]

\tikzset{threshold_fig/.pic={
\draw (-1,0)--(1,0);
\draw[red,line width=1.0 pt] (-1,-1)--(0,-1)--(0,1)--(1,1);
}}

\tikzset{
  bridging path/.initial=arc,
  bridging span/.initial=8pt,
  bridging gap/.initial=4pt,
  bridge/.style 2 args={
    spath/split at intersections with={#1}{#2},
    spath/insert gaps after
    components={#1}{\pgfkeysvalueof{/tikz/bridging span}},
    spath/join components upright
    with={#1}{\pgfkeysvalueof{/tikz/bridging path}},
    spath/split at intersections with={#2}{#1},
    spath/insert gaps after
    components={#2}{\pgfkeysvalueof{/tikz/bridging gap}},
  }
}

\tikz[overlay] \path[spath/save=arc] (0,0) arc[radius=1cm, start
    angle=180, delta angle=-180];

\path		
    %Encoder
    (0,0)coordinate[](m){} 
    node(Encoder)[blk, text width=1.6cm, right=2.75 of m]{Encoder}

    %Radar precoder
    coordinate[below=3.5 of m](theta){}
    node(Radar_precoder)[blkNN, text width=1.6cm] at (theta -| Encoder){Precoding}
    %Comm. precoder
    coordinate[below=2 of theta](varphi){}
    node(Comm_precoder)[blkNN, text width=1.6cm]at (varphi -| Encoder){Precoding}

    %Joint precoder
    node(Joint_precoder)[blk, minimum height=1.4cm, text width=1.2cm, right=3.5 of $(Radar_precoder)!0.5!(Comm_precoder)$]{ISAC precoder $\bmf(\eta,\phi)$} 

    %Radar CH
    node(Radar_CH)[blkCH, below right=-1 and 7 of Encoder]{Sensing Channel \\ $\Yr \sim \Pr(\Yr|\bmf, \bmx)$}
    %Comm CH
    node(Comm_CH)[blkCH, below = 2 of Radar_CH]{Comm. Channel \\ $\yc \sim \Pr(\yc|\bmf, \bmx)$}
    %Coordinate to help connecting to CHs
    coordinate[below left=0.5 and 0.5 of Radar_CH.west](temp_radar)
    coordinate[above left=0.5 and 1 of Comm_CH.west](temp_comm)

    %Radar estimation
    node(Radar_estimator)[blkNN, minimum height=1.5cm, below right=0.4 and 4.75 of Radar_CH.north]{Sensing Estimator}
    %Coordinate for 2nd input of radar estimator
    coordinate[below left=1.1 and 5.5 of Radar_estimator.west](input_radar)
    %Output coordinates for radar
    coordinate[above right=-0.2 and 3 of Radar_estimator](output_radar_1)
    coordinate[below right =-0.2 and 3 of Radar_estimator](output_radar_2)
    ($(output_radar_1)!0.5!(output_radar_2)$)coordinate(output_radar_3)
    %Thresholding block for probability
    % node(Threshold)[draw, minimum height=0.8cm, text width=0.8cm, right=0 of output_radar_1]{} %The coordinate output_radar_1 is needed just so that the outputs of Radar estimator are evenly spaced
    % %Output of thresholding
    % coordinate[right=1 of Threshold](output_threshold)

    %Communication decoder
    node(Comm_decoder)[blk] at (Comm_CH -| Radar_estimator){Comm. Decoder}

    %Coordinate for CSI
    coordinate[below=1 of Comm_CH](csi)
    (csi-|Comm_decoder)coordinate(csi_2) %This is just to center the text of CSI
    
;
%Input to encoder
\draw[->] (m)--node[above]{$\bmm \in \Mset^S$}(Encoder);

%Input to radar precoder
\draw[->] (theta)--node[above]{$\{\thetamin, \thetamax\}$}(Radar_precoder);
%Input to comm precoder
\draw[->] (varphi)--node[above]{$\{\varphimin, \varphimax\}$}(Comm_precoder);

%From individual precoders to joint precoder
\draw[->] (Radar_precoder)--node[above]{$\bmf_r \in \mathbb{C}^K$}(Radar_precoder -| Joint_precoder.west);
\draw[->] (Comm_precoder)--node[above]{$\bmf_c \in \mathbb{C}^K$}(Comm_precoder -| Joint_precoder.west);

%Encoder and Joint precoder to CHs
\draw[-] (Joint_precoder)--node[above]{$\bmf \in \mathbb{C}^K$}(Joint_precoder-|temp_radar); % This is so that the text is centered before the vertical line
\draw[->] (Joint_precoder)--(Joint_precoder -| Comm_CH.west);
\draw[->] (temp_radar)--(temp_radar -| Radar_CH.west);
\draw[->] (Encoder)--node[above]{$\bmx(\bmm) \in \mathbb{C}^S$}(Encoder -| Radar_CH.west);
\draw[-] (Encoder)-|(temp_comm);
% Draw arc (bridge) in intersection point
\path[spath/save=up] (Joint_precoder)-|(temp_radar);
\path[spath/save=along] (temp_comm)--(temp_comm -| Comm_CH.west);
\tikzset{bridge={along}{up}}
\draw[-,spath/use=up];
\draw[->,spath/use=along];

%Radar receiver
\draw[->] (Radar_CH)--node[above]{$\Yr \in \complexset[K][S]$}(Radar_CH-|Radar_estimator.west);
\draw[->] (input_radar)--node[above]{$\{\thetamin, \thetamax, \Rmin, \Rmax\}$}(input_radar-|Radar_estimator.west);
\draw[->] (Radar_estimator.east |- output_radar_1)--node[above]{$\that \in \{0,1\}$}(output_radar_1);
% \pic[scale=0.5]() at (Threshold) {threshold_fig};
% \draw[->] (Threshold)--node[above]{$\that$}(output_threshold);
\draw[->] (Radar_estimator.east |- output_radar_2)--node[above]{$\hat{R} \in \mathbb{R}_{\geq 0}$}(output_radar_2);
\draw[->] (Radar_estimator.east |- output_radar_3)--node[above]{$\hat{\theta} \in [-\frac{\pi}{2}, \frac{\pi}{2}]$}(output_radar_3);

%Comm receiver
\draw[->] (Comm_CH)--node[above]{$\yc \in \mathbb{C}^S$}(Comm_decoder);
\draw[->] (Comm_decoder)--node[above]{$\hat{\bmm} \in \Mset^S$}(Comm_decoder-|output_radar_1);

%CSI
\draw[-, dotted] (Comm_CH)--(csi);
\draw[-, dotted] (csi)--node[above]{$\bmkappa =  \bmbeta\bmf^\top\atx(\varphi)$}(csi_2);
\draw[->, dotted] (csi_2)--(Comm_decoder);

\end{tikzpicture}
    \caption{Block diagram of the ISAC system model. The colored blocks can be implemented following the baseline of Sec.~\ref{sec_baseline}, or model-based learning of Sec.~\ref{sec_mb_learning}. The precoding block applies the same mapping function for sensing and communication. Note that the sensing estimator is co-located with the ISAC transmitter.}
    \label{fig_system_model}
\end{figure*}

This section provides the mathematical description of the \ac{ISAC} transmitted signal, the received signals at the sensing and communication ends, and the hardware impairments. A block diagram of the system model is depicted in Fig.~\ref{fig_system_model}.

\subsection{Single-Target Sensing Model} \label{subsec_sensing_model}
We consider an \ac{ISAC} transceiver, equipped with a \ac{ULA} of $K$ antenna elements, which sends \ac{OFDM} signals with $S$ subcarriers and a subcarrier spacing of $\Delta_f$ Hz. The \ac{ISAC} transceiver senses a single target in the environment \textcolor{myred}{. Denoting $t=0$ as target absence and $t=1$ otherwise, we assume that $\Pr(t=0)=\Pr(t=1)=1/2$.}
The backscattered signal at the sensing receiver $\Yr\in\complexset[K][S]$ can be expressed in the spatial-frequency domain \cite{OFDM_radar_TVT_2020, MIMO_OFDM_ICI_JSTSP_2021, 5G_NR_JRC_analysis_JSAC_2022} as
\begin{align}
    \Yr = 
    \begin{cases}        1/\sqrt{S}\psi\arx(\theta)\atx^{\top}(\theta)\bmf[\bmx(\bmm) \odot \bmrho(\tau)]^{\top} + \bmW, & t=1 \\
        \bmW, & t=0,
    \end{cases} 
    \label{eq_Yr}
\end{align}
which represents a binary hypothesis testing problem. In \eqref{eq_Yr}, \textcolor{myred}{$(\cdot)^\top$ denotes the transpose operation, and} $\psi \sim \cnormal(0, \sigma_r^2)$ denotes the complex normal channel gain according to the Swerling-1 target model, with $\sigma_r^2$ representing path attenuation and radar cross section effects. The steering vectors of the receive and transmit \ac{ULA} are represented by $\arx, \atx \in \complexset[K]$, respectively. In the absence of hardware impairments, $[\arx(\theta)]_k = [\atx(\theta)]_k = \exp(-\jmath 2\pi (k-(K-1)/2)d \sin(\theta)/\lambda)$, $k=0,...,K-1$, with $d = \lambda / 2$, $\lambda = c/f_c$, where $c$ is the speed of light in vacuum and $f_c$ is the carrier frequency. The energy of the transmit \ac{ULA} is steered by the precoder $\bmf \in \complexset[K]$. The communication messages $\bmm \in \Mset^S$ are conveyed in $\bmx(\bmm) \in \complexset[S]$, where each message is uniformly drawn from the set $\Mset$. The range of the target induces a phase shift in the received \ac{OFDM} signal across subcarriers, which is expressed in $\bmrho(\tau) \in \complexset[S]$, such that $[\bmrho(\tau)]_s = \exp(-j2\pi s \Deltaf\tau)$, $s=0,...,S-1$, and where $\tau = 2R/c$ represents the round-trip time of a target $R$ meters away from the transmitter. The target angle and range are uniformly distributed as $\theta \sim \U[\thetamin, \thetamax]$, and $R \sim \U[\Rmin, \Rmax]$, respectively. The receiver noise is represented by $\bmW$, with $\vecc{\bmW}\sim\mathcolor{myred}{\cnormal}(\bm{0}, N_0\bmI_{KS})$. The integrated sensing \ac{SNR} across $K$ antenna elements is $\SNRr = K\sigma_r^2/N_0$.

The transmitter and the co-located receiver are assumed to have a coarse estimate of the target position, i.e., $\{\thetamin, \thetamax, \Rmin, \Rmax\}$ are known to the \ac{ISAC} transceiver. The goals of the receiver are: (i) \textcolor{myred}{minimize the probability of misdetection of the target $\Pmd = \Pr(\that=0|t=1)$ for a given false alarm $\Pfa = \Pr(\that=1|t=0)$}, where $\that$ is the estimate of the target presence, and (ii) minimize the position \ac{RMSE}, $\sqrt{\mathcolor{myred}{\mathbb{E}}\{\norm{\bmp-\bmphat}^2}\}$, where the target position is computed from its angle and range as
\begin{align}
    \bmp = \begin{bmatrix}
            R\cos{(\theta)} & R\sin{(\theta)}
           \end{bmatrix}^\top,
    \label{eq:pos_from_angle_range}
\end{align}
and $\bmphat$ is the estimated position.

\subsection{MISO Communication Model} \label{subsec_comm_model}
A single-antenna-element \ac{UE} is assumed to always receive the communication signal emitted by the \ac{ISAC} transceiver. The received signal $\yc\in\complexset[S]$ by the \ac{UE} is formulated in the frequency domain as \cite{overview_SP_JCS_JSTSP_2021}
\begin{align}
    \yc = [\bmx(\bmm)\odot \bmbeta]\bmf^\top\atx(\varphi) + \bmn, \label{eq_yc}
\end{align}
with $\bmbeta = \bmF[\beta_0, ..., \beta_{L-1}, 0, ..., 0]$, where $\bmF\in\complexset[S][S]$ is the unitary \ac{DFT} matrix, and $L$ is the number of channel taps. Each channel tap is distributed according to $\beta_l \sim \cnormal(0,\sigma_l^2)$. The angle of the \ac{UE} is distributed as $\varphi\sim\U[\varphimin, \varphimax]$. The communication receiver noise follows $\bmn \sim \cnormal(\bm{0}, N_0\bmI_S)$. The average communication \ac{SNR} across subcarriers is $\SNRc= \sum_{l=1}^L\sigma_l^2/(SN_0)$.
The \ac{ISAC} transceiver is also considered to know $\{\varphimin, \varphimax\}$. The goal of the \ac{UE} is to minimize the \ac{SER}, $\Pr([\bmmhat]_s\neq[\bmm]_s)$, with $[\bmmhat]_s$ the estimated symbol of the $s$-th subcarrier. To this end, the \ac{UE} is fed with the \ac{CSI} $\bmkappa =  \bmbeta\bmf^\top\atx(\varphi)$.

\subsection{ISAC Transmitter} \label{subsec_isac_transmitter}
To study the achievable \ac{ISAC} trade-offs of the considered system, we design a flexible sensing-communication beamformer $\bmf \in \complexset[K]$, based on a sensing precoder $\bmf_r \in \mathbb{C}^K$, and a communication precoder $\bmf_c\in \mathbb{C}^K$, following the multi-beam approach of \cite{zhang2018multibeam} as
\begin{align} \label{eq_isac_precoder_bench}
    \bmf(\eta,\phi) = \frac{\sqrt{\eta} \bmf_r + \sqrt{1-\eta}e^{\jmath \phi }\bmf_c}{\Vert\sqrt{\eta} \bmf_r + \sqrt{1-\eta}e^{\jmath \phi }\bmf_c \Vert },
\end{align}
where $\eta \in [0,1]$ is the ISAC trade-off parameter, and $\phi \in [0,2 \pi)$ is a phase ensuring coherency between multiple beams. By varying $\eta$ and $\phi$, the \ac{ISAC} trade-offs are explored. The sensing precoder $\bmf_r$ points to the angular sector of the target, $[\thetamin, \thetamax]$, whereas the communication precoder $\bmf_c$ points to the angular sector of the \ac{UE}, $[\varphimin, \varphimax]$. Secs.~\ref{subsec_baseline_beamformer} and \ref{subsec_mb_beamformer} detail how the baseline and \ac{MB-ML} design the precoder for an angular sector, respectively. This operation is identically applied to obtain $\bmf_r$ and $\bmf_c$, as depicted in Fig.~\ref{fig_system_model}.

\subsection{Hardware Impairments} \label{subsec_hardware_impairments}
As represented in Fig.~\ref{fig_scenario}, we consider antenna impairments in the \ac{ULA} of the \ac{ISAC} transceiver, which affect the steering vector models in \eqref{eq_Yr} and \eqref{eq_yc}. Impairments in the antenna array include mutual coupling, array gain errors, or antenna displacement errors, among others \cite{schenk2008rf}. Following the impairment models of \cite{chen2023modeling}, our steering vector model is conditioned on an unknown perturbation vector $\bmd$, where the meaning and dimensionality of $\bmd$ depend on the type of impairment. We thus write the perturbed steering vector model as $\aperturbed(\theta;\bmd)$.

\section{Baseline} \label{sec_baseline}
This section describes the baseline algorithms to obtain the transmit \ac{ISAC} precoder, and to process the received sensing and communication signals. The baseline is rooted in model-based benchmarks, which assume perfect knowledge of the system model of Sec.~\ref{sec_model}. The description of the baseline is included since MB-ML in Sec.~\ref{sec_mb_learning} uses a variation of the algorithms described here. The baseline is later compared with \ac{MB-ML} in Sec.~\ref{sec_results}.

\subsection{Beamformer} \label{subsec_baseline_beamformer}
We first design a precoder that points towards a given angular sector, which yields the individual sensing ($\bmf_r$) and communication ($\bmf_c$) precoders to be later combined in \eqref{eq_isac_precoder_bench}. We resort to the beampattern synthesis approach in \cite{precoding_mmWave_JSTSP_2014,analogBeamformerDesign_TSP_2017}, as follows: define a uniform angular grid covering $[-\pi/2, \pi/2]$ with $\Ntheta$ grid locations $\{\vartheta_i\}_{i=1}^{\Ntheta}$. Given an angular interval $\bmvarthetainterval = [\varthetamin, \varthetamax]$, we denote by $\bbold \in \complexset[\Ntheta]$ the desired beampattern over the defined angular grid, which follows
\begin{align}
    [\bbold]_i = 
    \begin{cases}
        K, &~~ \text{if} ~~\vartheta_i \in \bm{\vartheta}_{\text{interval}} \\
        0, &~~ \text{otherwise.} 
    \end{cases} \label{eq_desired_beampattern}
\end{align}
The beampattern synthesis problem can then be formulated as 
$\mathop{\mathrm{min}}\limits_{\bmfbs} \lVert \bbold - \bmAtx^\top \bmfbs  \rVert_{2}^2$, where $\bmAtx = [\atx(\vartheta_1) \, \ldots \, \atx(\vartheta_{\Ntheta})] \in \complexset[K][\Ntheta]$ denotes the transmit steering matrix evaluated at the grid locations. This \ac{LS} problem
has a simple closed-form solution 
\begin{equation}\label{eq_LS_precoder_baseline}
    \bmfbs = (\bmAtx^\conj \bmAtx^\top)^{-1} \bmAtx^\conj \bbold,
\end{equation}
\textcolor{myred}{with $(\cdot)^{-1}$, and $(\cdot)^\conj$ the inverse and complex conjugate operations, respectively.} Depending on the input angular interval, we obtain $\bmf_r$ or $\bmf_c$ as the result of \eqref{eq_LS_precoder_baseline}, as depicted in Fig.~\ref{fig_system_model}.

\subsection{Sensing Receiver} \label{subsec_baseline_sensing_rx}
In the considered monostatic sensing setup, the receiver has access to communication data $\bmx(\bmm)$, which enables removing its impact on the received signal \eqref{eq_Yr} via reciprocal filtering \cite{OFDM_Radar_Corr_TAES_2020,reciprocalFilter_OFDM_2023}
\begin{align}\label{eq_Yr_tilde}
    \Yrtilde = \Yr \oslash \boldone \bmx^\top(\bmm) =  
    \begin{cases}
        \alpha \arx(\theta) \bmrho^\top(\tau) + \bmN,&~~  t=1   \\
        \bmN,&~~ t=0,
    \end{cases}    
\end{align}
where $\alpha=\atx^\top(\theta) \bmf \psi / \sqrt{S}$ and $\bmN = \bm{W} \oslash \boldone \bmx^\top(\bmm)$.

The radar detection problem in \eqref{eq_Yr_tilde} involves random parameters $\alpha$, $\theta$ and $\tau$. Hence, we resort to the maximum a-posteriori (MAP) ratio test (MAPRT) detector \cite{MAP_Detector_TSP_2021} as our detector benchmark, which takes into account the prior information on $\alpha$, $\theta$ and $\tau$. Let $\hypz$ and $\hypone$ denote the absence and the presence of a target, respectively, in \eqref{eq_Yr_tilde}. Then, the corresponding MAPRT is given by \cite{MAP_Detector_TSP_2021}
\begin{align}\label{eq_maprt}
    \Lambda(\Yrtilde) = \frac{ \max_{\alpha, \theta, \tau} p(\alpha, \theta, \tau, \hypone \, \lvert \, \Yrtilde ) }{ p(\hypz \, \lvert \, \Yrtilde ) } \hdet \mathcolor{myred}{\nubar} ~,
\end{align}
\textcolor{myred}{where $\nubar$ is the detection threshold that can be determined based on a preset false alarm probability \cite[Ch.~7]{richards2005fundamentals}}. We assume that $p(\hypz) = p(\hypone) = 1/2$ and $\bmx(\bmm)$ are PSK symbols\footnote{It can be readily shown that the same detector in \eqref{eq_maprt_etabar_no_y} is obtained also for arbitrary constellations by following \cite[App.~A]{mateos2022end}.}, keeping noise statistics the same, i.e., $\vecc{\bmN}\sim\normal(\bm{0}, N_0\bmI_{KS})$. Following similar steps to those in \cite[App.~A]{mateos2022end}, \eqref{eq_maprt} becomes
\begin{align}\label{eq_maprt_etabar_no_y}
    \absbigs{ \arx^\hermit(\thetahat) \Yrtilde \bmrho^\conj(\tauhat)  }   \hdet \nu ~,
\end{align}
\textcolor{myred}{where $\nu \propto \nubar$} and
\begin{align}\label{eq_maprt_etabar_no_y2}
    (\thetahat, \tauhat) = \arg \max_{\substack{\theta \in [\thetamin, \thetamax] \\  \tau \in [\taumin, \taumax]}} ~ \absbigs{ \arx^\hermit(\theta) \Yrtilde \bmrho^\conj(\tau)  }   ~.
\end{align}
Here, $\taumin = 2\Rmin/c$ and $\taumax = 2\Rmax/c$.

\subsection{Communication Receiver} \label{subsec_baseline_comm_rx}
Assuming that the \ac{UE} receives the \ac{CSI} $\bmkappa =  \bmbeta\bmf^\top\atx(\varphi)$ as shown in Fig.~\ref{fig_system_model}, we perform maximum likelihood estimation of the communication symbols as in \cite{mateos2022model, mateos2023model}. 

\section{Model-based Learning} \label{sec_mb_learning}
The goal of \ac{MB-ML} is to perform successful target detection and position estimation (according to the performance metrics defined in Sec.~\ref{sec_model}), while simultaneously learning the perturbation vector $\bmd$ of the steering vector $\aperturbed(\theta;\bmd)$. This section describes the details of the beamforming and sensing estimation algorithms tailored to \ac{MB-ML}, which are largely rooted in the baseline operations introduced in Sec.~\ref{sec_baseline}. We perform end-to-end learning of both transmitter and receiver, as shown in Fig.~\ref{fig_system_model}. The MB-ML method considered in this paper is a special case of the multi-target approach proposed in \cite[Sec. IV]{mateos2023model}, applied to a single sensing target. We also describe the supervised and unsupervised loss functions to perform SL, UL, and SSL of the \ac{MB-ML} model. 

\subsection{Beamformer} \label{subsec_mb_beamformer}
\ac{MB-ML} beamforming performs the same operations as \eqref{eq_desired_beampattern} and \eqref{eq_LS_precoder_baseline}. However, the transmit steering matrix $\bmAtx$ involves perfect knowledge of the steering vector model and hence the perturbation vector $\bmd$. To compensate for unknown antenna impairments, \ac{MB-ML} instead constructs a new steering matrix $\Aperturbed(\bmd) = [\aperturbed(\theta_1;\bmd)\,\ldots\,\aperturbed(\theta_{\Ntheta};\bmd)]$, which is used in \eqref{eq_LS_precoder_baseline} to obtain a sensing precoder $\bmf_r$, or a communication precoder $\bmf_c$.

\subsection{Sensing Receiver} \label{subsec_mb_sensing_rx}
As a first step, \ac{MB-ML} removes the effect of the communication symbols in the received signal following \eqref{eq_Yr_tilde}.
However, \ac{MB-ML} cannot directly apply the same operations \eqref{eq_maprt_etabar_no_y}-\eqref{eq_maprt_etabar_no_y2} as the baseline sensing receiver of Sec.~\ref{subsec_baseline_sensing_rx}. The nondifferentiability of the $\arg\max$ operation in \eqref{eq_maprt_etabar_no_y2} impedes gradient flow during backpropagation. Algorithm~\ref{alg_mb_sensing_rx} describes the operations performed in the \ac{MB-ML} sensing estimator, where $[\bm{A}]_{n:m, l:p}$ denotes the submatrix of $\bm{A}$ covering from the $n$-th to the $m$-th row and from the $l$-th to the $p$-th column of $\bm{A}$. The main differences with the receiver baseline are:
\begin{enumerate}
    \item \textbf{Select a window of elements around the maximum of the angle-delay map:} We leverage on the angle and range resolution of the sensing system to select the elements of the angle-delay map that correspond to the target. The angle and range resolutions in our case are
    \begin{align}
        \Delta_{\theta} [\text{rad}] \approx \frac{2}{K}, \quad \Delta_R [\text{m}] \approx \frac{c}{2S\Deltaf}.
    \end{align}   
    The number of elements to select in the angle-delay map is then given by
    %which after normalization as
    \begin{align} \label{eq_delta_resolutions}
        \deltatheta = \bigg \lfloor \frac{\Delta_{\theta}\Ntheta}{\thetamax-\thetamin}\bigg \rfloor, \quad \deltaR =  \bigg \lfloor\frac{\Delta_R N_{\tau}}{\Rmax-\Rmin} \bigg \rfloor .
    \end{align} 
    Invalid elements in the selected window, e.g., negative rows or columns, are discarded. This approach is especially useful in a multi-target scenario, which was implemented in \cite{mateos2023model}. Moreover, this approach is similar to \emph{hard thresholding} techniques, in which selecting a subset of elements from the data makes learning faster \cite{yassine2022}.
    \item \textbf{Softmax operation:} We apply the softmax operator to the selected elements in the angle-delay map. This differentiable operation allows to obtain an estimate of the probability that each element in the angle-delay map corresponds to the target. From this probability matrix, we can estimate the position of the target by a weighted average of considered angles and ranges, as described in Algorithm~\ref{alg_mb_sensing_rx}.
\end{enumerate}
Note that although we perform a nondifferentiable operation in \eqref{eq_argmax_mbml}, the result is only used to slice the angle-delay map $\admap$, performing differentiable operations from that point on.

\begin{algorithm}[tb]
    \caption{Model-based learning sensing estimation.}
    \label{alg_mb_sensing_rx}
    \begin{algorithmic}[1]
        \State \textbf{Input:} Observation $\Yrtilde$ in \eqref{eq_Yr_tilde}, grid vectors $\thetagrid$ and $\taugrid$, discrete angle and range resolutions $\deltatheta, \deltaR$ in \eqref{eq_delta_resolutions}, perturbed impairment vector $\bmd$, detection threshold $\eta$.
        \State \textbf{Output:} Estimate of the target presence $\that$ and its position, $\bmphat$, if applicable.
        \State Construct the dictionaries 
        \begin{align}
        \bmPhi_{\rm{pert}} &= [\aperturbed(\theta_1;\bmd)\,\ldots\,\aperturbed(\theta_{\Ntheta};\bmd)], \\ \bmPhi_d &= [\bmrho(\tau_1)\,\ldots\,\bmrho(\tau_{N_{\tau}})].
        \end{align}
        \State Compute the angle-delay map 
        \begin{align}\label{eq_admap}
            \admap = \absbigs{\bmPhi_{\rm{pert}}^H \Yrtilde \bmPhi_d^\conj  }.
        \end{align}
        \State \textbf{if} $\max_{i,j} \admapindex{i}{j} > \eta$
        \Indent 
        \State Matrix element that maximizes the angle-delay map:
            \begin{align} \label{eq_argmax_mbml}
                 (\ihat, \jhat) = \arg \max_{i,j} \admapindex{i}{j} ~.
            \end{align}
        \State Select $\deltatheta$ rows and $\deltaR$ columns around $(\ihat, \jhat)$:
            \begin{align}
                \admapreduced = \admapindex{\ihat-\deltatheta:\ihat+\deltatheta}{\jhat-\deltaR:\jhat+\deltaR} ~.
            \end{align}
        \State Apply softmax: $\admapreducedsoft = \rm{Softmax}(\admapreduced)$.
        \State Angle-delay estimation by weighted average:
        \begin{align}
            \thetahat &= \sum_{n=1}^{2\deltatheta+1} \theta_{n+\ihat-\deltatheta-1} \sum_{m=1}^{2\deltaR+1} [\admapreducedsoft]_{n,m}~,\\
            \Rhat &= \sum_{m=1}^{2\deltaR+1} R_{m+\jhat-\deltaR-1} \sum_{n=1}^{2\deltatheta+1} [\admapreducedsoft]_{n,m}~.
        \end{align}
        \State $\bmphat \gets [\Rhat\cos(\thetahat)\, \Rhat\sin(\thetahat)]^\top$.
        \State $\that \gets 1$.
        \EndIndent
        \State \textbf{else}
        \Indent
        \State $\that \gets 0$. \Comment{No position is estimated.}
        \EndIndent
    \end{algorithmic}
\end{algorithm}

\subsection{Loss Functions} \label{subsec_loss_functions}
One of the goals introduced in Sec.~\ref{sec_introduction} is to perform a comparison between SL, UL, and SSL. Here we introduce the corresponding supervised and unsupervised loss functions, which leverage labeled and unlabeled data, respectively.

\subsubsection{Supervised Loss Function} \label{subsubsec_loss_supervised}
For SL, the labeled data contains the true position of a present target in the environment. The \ac{MB-ML} model is then trained based on the \ac{MSE} loss between the true and estimated positions, as
\begin{align} \label{eq_loss_super}
    \Jsuper = \E \big[ \norm{\bmp-\bmphat}^2 \big].
\end{align}
\subsubsection{Unsupervised Loss Function} \label{subsubsec_loss_unsupervised}
Inspired by the baseline sensing receiver, the operation in \eqref{eq_maprt_etabar_no_y2} accounts for the fact that at the true target position, the steering vector model is matched to the received signal, maximizing the angle-delay map. Nevertheless, under a mismatched steering vector model, the angle-delay map is no longer maximized. Hence, we propose to use as unsupervised training loss the negative maximum of the angle-delay map of the received signal, i.e.,
\begin{align} \label{eq_loss_unsuper}
    \Junsuper = -\max_{i,j} \admapindex{i}{j}.
\end{align}
Minimizing \eqref{eq_loss_unsuper} ensures convergence of the perturbation vector $\bmd$ to its true value by seeking optimal phase alignment with the ULA steering vector in \eqref{eq_admap}.

\subsubsection{Semi-supervised Learning} \label{subsubsec_loss_semisupervised}
We perform SSL in a sequential manner, i.e., we first perform SL or UL, and then switch to the other. 
The order of SSL is investigated in Sec.~\ref{sec_results}, where we also explore how the performance of SSL evolves as a function of the amount of labeled data.

\section{Results} \label{sec_results}
This section describes the simulation parameters and the results\footnote{Source code to reproduce all numerical results in this paper is available at \url{github.com/josemateosramos/SSLISAC}.}, comparing (i) the baseline of Sec.~\ref{sec_baseline}, (ii) \ac{SL} relying on labeled data, (iii) \ac{UL}, where no labeled data is utilized, and (iv) \ac{SSL}, which uses a mix of labeled and unlabeled data.

\subsection{Simulation Parameters}  \label{subsec_parameters}
We consider a \ac{ULA} composed of $K=64$ antenna elements. The \ac{OFDM} signal includes $S=256$ subcarriers, with a subcarrier spacing of $\Delta_f = 120$~kHz, and a carrier frequency of $f_c=60$~GHz. The cardinality of the set of messages is $|\Mset| = 4$, and the communication encoder is set as a \ac{QPSK} modulator. The number of channel taps in the communication channel is set to $L=5$ taps, with an exponential power delay profile, i.e., $\sigma_l^2 = \exp{(-l)}, l=0,...,L-1$. The channel taps are normalized to give a certain communication \ac{SNR}. The sensing and communication \acp{SNR} are $\SNRr = 15$~dB and $\SNRc = 20$~dB, respectively. We adopt the impairment model of \cite{mateos2022end}, i.e., we consider inter-antenna spacing impairments, such that $\bmd \sim \cnormal((\lambda/2)\boldone, \sigma_{\lambda}^2 \bmI_K)$, and the perturbed steering vector model becomes $[\aperturbed(\theta; \bmd)]_k = \exp(-\jmath 2 \pi (k-(K-1)/2) [\bmd]_k \sin (\theta ) / \lambda)$. We select a standard deviation of $\sigmalambda = \lambda/25 = 0.2$ mm. The initial perturbation vector for \ac{MB-ML} is $\bmd = (\lambda/2) \boldone$.

We train \ac{MB-ML} for a wide variety of target angular sectors, i.e., we randomly draw $\{\thetamin, \thetamax\}$ in each transmission as in \cite{mateos2023model}. We consider $\thetamean \sim \U[-60\degree, 60\degree]$ and $\thetaspan \sim \U[10\degree, 20\degree]$, from which we compute $\{\thetamin, \thetamax\} = \{\thetamean-\thetaspan/2, \thetamean+\thetaspan/2\}$. Conversely, the target range sector is fixed for all transmissions as $[\Rmin, \Rmax] = [0,200]$~m. The communication angular sector is set to $[\varphimin, \varphimax] = [30\degree, 50\degree]$. To construct the angle and range dictionaries $\bmPhi_{\text{pert}}$ and $\bmPhi_d$ for \ac{MB-ML}, we consider a grid of angles $\thetagrid$ covering $[-\pi/2, \pi/2]$ and a grid of delays $\taugrid$ covering the range interval $[0,200]$~m, which are also used to compute the discrete angle and range resolutions in \eqref{eq_delta_resolutions}. The grid size for angle and range are $\Ntheta=720$ and $N_{\tau} = 200$, respectively. We use the Adam optimizer \cite{kingma2015adam}. To tune the hyperparameters of SL and UL, we tested learning rates ranging from $10^{-8}$ to $10^{-7}$ with an increment of $10^{-8}$, batch sizes of 3000 and 4500 samples, and we reduced the learning rate when the loss function plateaued\footnote{Higher learning rates induce divergence of the loss function since the true inter-antenna spacing $\bmd$ is in the order of $\lambda$/2.}. The best performance was obtained with a learning rate of $4\cdot 10^{-7}$ for SL and $5\cdot 10^{-7}$ for UL, along with a batch size of 3000 samples. Additionally, we observed that for SL, reducing the learning rate to $4\cdot10^{-8}$ after 50,000 iterations yielded better results. To test the sensing performance, we fixed the angular sector to $[\thetamin, \thetamax] = [-40\degree, -20\degree]$, although the conclusions of the upcoming sections also apply to other testing angular sectors. The ISAC precoder of \eqref{eq_isac_precoder_bench} is computed by varying $\rho$ and $\phi$ in the intervals $[0,1]$ and $[0, 14\pi/8]$, respectively, with 8 grid points each.

\subsection{Sensing Results} \label{subsec_results_sensing}
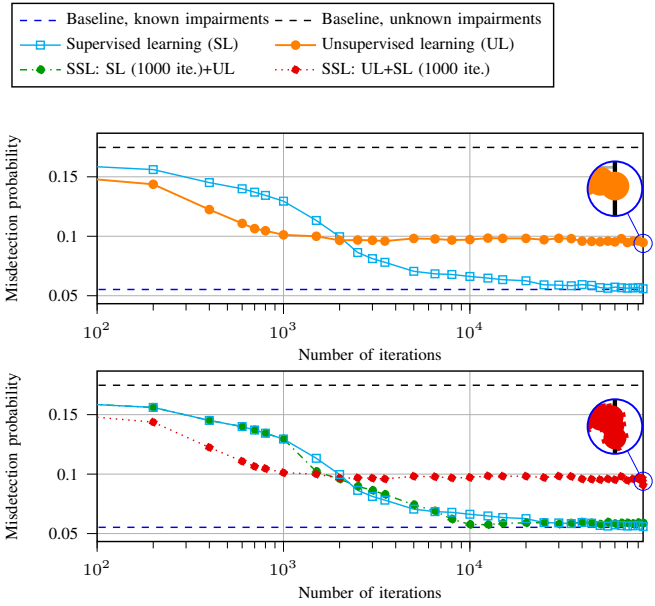
\begin{figure}[tb]
    \centering
    \begin{tikzpicture}[font=\scriptsize, spy using outlines=
	{circle, magnification=3, connect spies}]

%%%%%%%%%%%%%%%%%%%%%%%%%%%%%%%%%%%%%%%%%%%%%%%%
% First plot, iterations vs Misdetection Prob. %
%%%%%%%%%%%%%%%%%%%%%%%%%%%%%%%%%%%%%%%%%%%%%%%%
\begin{axis}[
width=8cm,
height=2.5cm,
scale only axis,
legend cell align={left},
legend columns = 2,
legend style={
  fill opacity=1,
  draw opacity=1,
  text opacity=1,
  at={($(current axis.west)+(-0.84,2.75cm)$)}, % Adjust the y-coordinate to move the legend above the plot
  anchor=west
},
legend entries={
Baseline, known impairments \\
Baseline, unknown impairments \\
Supervised learning (SL) \\
Unsupervised learning (UL) \\
SSL: SL (1000 ite.)+UL \\
SSL: UL+SL (1000 ite.) \\
},
log basis y={10},
tick align=outside,
tick pos=left,
x grid style={white!69.0196078431373!black},
xlabel={Number of iterations},
xmajorgrids,
xmin=100, xmax=8.5e4,
xmode=log,
xtick style={color=black},
y grid style={white!69.0196078431373!black},
ylabel={Misdetection probability},
ymajorgrids,
ytick style={color=black},
ytick={0.05, 0.1, 0.15},
yticklabels={0.05, 0.1, 0.15},
]
\addplot[color=blue!90!black, dashed, mark options={scale=0.8}] table {result_figures/data_iterations/known_it_pmd.txt};

\addplot[color=black, dashed, mark options={scale=0.8}] table {result_figures/data_iterations/unknown_it_pmd.txt};

\addplot[color=cyan, mark=square, mark options={scale=0.8}] table {result_figures/data_iterations/super_it_pmd.txt};

\addplot[thick, color=orange, solid, mark=*, mark options={scale=0.8,solid }] table {result_figures/data_iterations/unsuper_it_pmd.txt};

\addlegendimage{color=green!60!black, mark=*, dashdotted, mark options={scale=0.8}}

\addlegendimage{color=red!90!black, mark=*, dotted, mark options={scale=0.8}}

\addlegendimage{color=magenta, mark=*, dashdotted, mark options={scale=0.8}}

% %Add magnifying coordinates
\coordinate (spypoint) at (axis cs:85000,0.094);
\coordinate (magnifyglass) at (axis cs:6e4,0.14);

\end{axis}

\spy [blue, size=0.8cm] on (spypoint) in node[fill=white] at (magnifyglass);

% %%%%%%%%%%%%%%%%%%%%%%%%%%%%%%%%%%%%%%%%%
% % Second plot %
% %%%%%%%%%%%%%%%%%%%%%%%%%%%%%%%%%%%%%%%%%

\begin{axis}[
width=8cm,
height=2.5cm,
at={(0,-200)},      %<- Specify position
scale only axis,
legend cell align={left},
log basis y={10},
tick align=outside,
tick pos=left,
x grid style={white!69.0196078431373!black},
xlabel={Number of iterations},
xmajorgrids,
xmin=100, xmax=8.5e4,
xmode=log,
xtick style={color=black},
y grid style={white!69.0196078431373!black},
ylabel={Misdetection probability},
ymajorgrids,
ytick style={color=black},
ytick={0.05, 0.1, 0.15},
yticklabels={0.05, 0.1, 0.15},
]

\addplot[color=blue!90!black, dashed, mark options={scale=0.8}] table {result_figures/data_iterations/known_it_pmd.txt};

\addplot[color=black, dashed, mark options={scale=0.8}] table {result_figures/data_iterations/unknown_it_pmd.txt};

\addplot[color=green!60!black, mark=*, dashdotted, mark options={scale=0.8}] table {result_figures/data_iterations/semisuper_1000_it_pmd.txt};

\addplot[color=red!90!black, mark=*, dotted, mark options={scale=0.8}] table {result_figures/data_iterations/semiunsuper_1000_it_pmd.txt};

\addplot[color=cyan, mark=square, mark options={scale=0.8}] table {result_figures/data_iterations/super_it_pmd.txt};

% %Add magnifying coordinates
\coordinate (spypoint_2) at (axis cs:85000,0.094);
\coordinate (magnifyglass_2) at (axis cs:6e4,0.14);

\end{axis}

\spy [blue, size=0.8cm] on (spypoint_2) in node[fill=white] at (magnifyglass_2);

\end{tikzpicture}
    \caption{Misdetection probability as a function of the number of training iterations, under hardware impairments. The false alarm probability was fixed to $\Pfa=10^{-2}$. In the baseline case, there is no training procedure. SSL: semi-supervised learning.}
    \label{fig_num_it_vs_performance}
\end{figure}

In the following, we use misdetection probability to assess sensing performance, although it was verified that the same conclusions hold when applying position RMSE as a performance metric.
Fig.~\ref{fig_num_it_vs_performance} shows the misdetection probability $\Pmd$ as a function of the number of training iterations for a fixed false alarm probability of $\Pfa=10^{-2}$, where the threshold $\eta$ in \eqref{eq_maprt_etabar_no_y} to yield a fixed $\Pfa$ was found empirically. 
We maintain a consistent batch size across all learning approaches to ensure a fair comparison. The performance of the baseline (which does not require any training) is shown as a reference.  Fig.~\ref{fig_num_it_vs_performance} (top) shows that UL initially converges faster than SL, but SL outperforms UL after around 20,000 iterations and achieves better final performance due to the use of ground-truth information. Fig.~\ref{fig_num_it_vs_performance} (bottom) compares SL with two SSL approaches: (i) starting with SL and switching to UL (SL+UL), and (ii) starting with UL and switching to SL (UL+SL). In both cases, the same total number of iterations as SL is used. The results indicate that starting with SL achieves better performance than starting with UL. 
Hence, from now on, we assume that \ac{SSL} first applies \ac{SL} followed by \ac{UL}. 

\begin{figure}[tb]
    \centering
    \begin{tikzpicture}[font=\small]

%%%%%%%%%%%%%%%%%%%%%%%%%
% First plot, ROC curve %
%%%%%%%%%%%%%%%%%%%%%%%%%
\begin{axis}[
width=10cm,
height=2.5cm,
scale only axis,
legend cell align={left},
log basis y={10},
tick align=outside,
tick pos=left,
x grid style={white!69.0196078431373!black},
xlabel={Labeled data ratio},
xmajorgrids,
xmin=5e-3, xmax=1,
xtick style={color=black},
y grid style={white!69.0196078431373!black},
ylabel={Misdetect. probability},
ymajorgrids,
xmode=log,
ytick style={color=black},
ytick={0.06, 0.08, 0.1},
yticklabels={0.06, 0.08, 0.1},
]

\addplot[color=cyan, only marks, mark=square, mark options={scale=1.2}] coordinates {(1,0.05584782508632791)};
\addlegendentry{Supervised learning (SL)};

\addplot[color=green!60!black, mark=*, mark options={scale=0.8}] table {result_figures/data_ratio/ratio_8_5e4.txt};
\addlegendentry{Semi-supervised learning (SSL)};

\addplot[color=cyan, dashed] coordinates {(1e-4,0.05584782508632791)  (1,0.05584782508632791)};

\end{axis}

\end{tikzpicture}
    \caption{Misdetection probability as a function of the labeled data ratio. The false alarm probability was fixed to $\Pfa=10^{-2}$ and the total number of iterations to 85,000.}
    \label{fig_pmd_vs_ratio}
\end{figure}
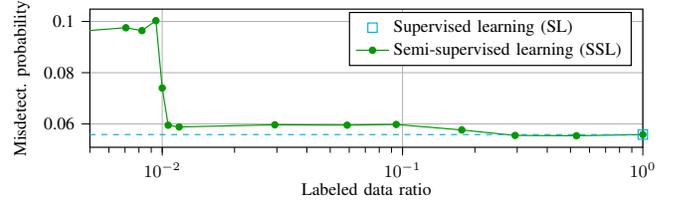

Fig.~\ref{fig_num_it_vs_performance} suggests that SSL with 1,000 SL iterations attains a comparable performance to fully SL, which is further investigated in the following.
Fig.~\ref{fig_pmd_vs_ratio} shows $\Pmd$ as a function of the labeled data ratio, defined as the proportion of labeled data to the total amount of data. SL, by definition, only uses labeled data and therefore has a ratio of 1 (a dashed line is included at the same performance level as SL for reference). Fig.~\ref{fig_pmd_vs_ratio} shows how SSL approaches the performance of SL with increasing amounts of labeled data, where SSL attains similar performance to SL with $1.2\%$ of the labeled data. 
This illustrates the potential of SSL to reduce the required labeled data and still obtain similar performance to SL.\footnote{We note that, SSL slightly outperforms SL in Fig.~\ref{fig_num_it_vs_performance} at higher labeled data ratios close to 1. 
However, we conjecture that this is an artifact of the fixed hyperparameter choices for SL described in Sec.~\ref{subsec_parameters}. As the size of the training data set tends to infinity, we expect SL to always perform at least as good as SSL assuming properly tuned hyperparameters.} However, below a certain cut-off amount of labeled data (approximately $1\%$ in Fig.~\ref{fig_pmd_vs_ratio}), SSL yields considerably worse performance than SL. This suggests that UL provides similar results to SL only with an initial point relatively close to the optimal solution. The cut-off value to obtain relevant SSL performance may depend on the sensing SNR. In low-SNR regimes for instance, the proposed unsupervised loss in \eqref{eq_loss_unsuper} may select spurious peaks due to noise, which would hinder UL and increase the cut-off value.

\subsection{ISAC Results} \label{subsec_isac_results}
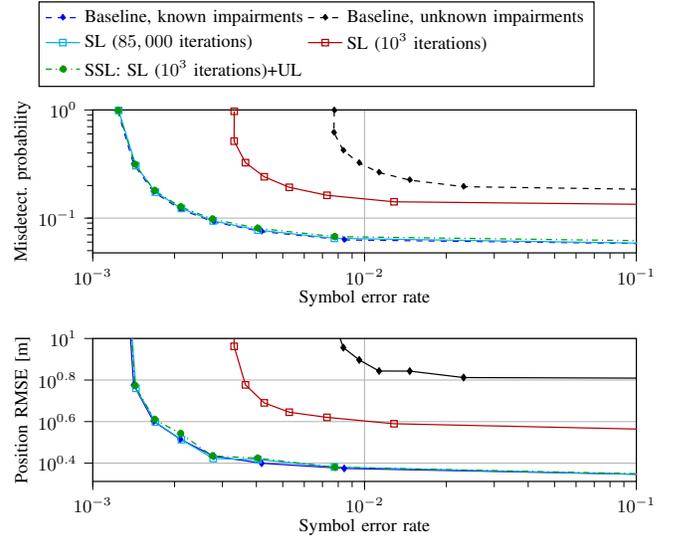
\begin{figure}[tb]
    \centering
    \begin{tikzpicture}[font=\small]

%%%%%%%%%%%%%%%%%%%%%%%%%%%%%%%%%%%%%%%%%%%%%%%
% First plot, SER vs misdetection probability %
%%%%%%%%%%%%%%%%%%%%%%%%%%%%%%%%%%%%%%%%%%%%%%%
\begin{axis}[
width=9.5cm,
height=2.5cm,
scale only axis,
legend cell align={left},
legend columns = 2,
legend style={
  fill opacity=1,
  draw opacity=1,
  text opacity=1,
  at={($(current axis)+(0.9,4.9cm)$)}, % Adjust the y-coordinate to move the legend above the plot
  anchor=west
},
log basis y={10},
tick align=outside,
tick pos=left,
x grid style={white!69.0196078431373!black},
xlabel={Symbol error rate},
xmajorgrids,
xmin=1e-3, xmax=1e-1,
xtick style={color=black},
y grid style={white!69.0196078431373!black},
ylabel={Misdetect. probability},
ymajorgrids,
ymax=1,
xmode=log,
ymode=log,
ytick style={color=black},
]

\addplot[color=blue!90!black, mark=diamond*, dashed, mark options={scale=0.8}] table {result_figures/data_isac/known_ser_pmd.txt};
\addlegendentry{Baseline, known impairments};

\addplot[color=black, mark=diamond*, dashed, mark options={scale=0.8}] table {result_figures/data_isac/unknown_ser_pmd.txt};
\addlegendentry{Baseline, unknown impairments};

\addplot[color=cyan, mark=square, mark options={scale=0.8}] table {result_figures/data_isac/super_ser_pmd.txt};
\addlegendentry{SL ($85,000$ iterations)};

\addplot[color=red!70!black, mark=square, mark options={scale=0.8}] table {result_figures/data_isac/super_1e3_ser_pmd.txt};
\addlegendentry{SL ($10^3$ iterations)};

\addplot[color=green!60!black, mark=*, dashdotted, mark options={scale=0.8}] table {result_figures/data_isac/ssl_1e3_ser_pmd.txt};
\addlegendentry{SSL: SL ($10^3$ iterations)+UL};

\end{axis}

%%%%%%%%%%%%%%%%%%%%%%%%%%%%%%%%%
% Second plot, Pfa vs Pos. RMSE %
%%%%%%%%%%%%%%%%%%%%%%%%%%%%%%%%%

\begin{axis}[
width=9.5cm,
scale only axis,
height=2.5cm,
at={(0,-1.1)},      %<- Specify position
log basis y={10},
tick align=outside,
tick pos=left,
legend columns = 2,
legend style={
  fill opacity=1,
  draw opacity=1,
  text opacity=1,
  at={(0.5,1.45)},
  anchor=center
},
x grid style={white!69.0196078431373!black},
xlabel={Symbol error rate},
xmajorgrids,
xmin=1e-3, xmax=1e-1,
xtick style={color=black},
y grid style={white!69.0196078431373!black},
ylabel={Position RMSE [m]},
ymajorgrids,
ymax=1e1,
xmode=log,
ymode=log,
ytick style={color=black},
]
\addplot[color=blue!90!black, mark=diamond*, mark options={scale=0.8}] table {result_figures/data_isac/known_ser_rmse_pos.txt};

\addplot[color=black, mark=diamond*, mark options={scale=0.8}] table {result_figures/data_isac/unknown_ser_rmse_pos.txt};

\addplot[color=cyan, mark=square, mark options={scale=0.8}] table {result_figures/data_isac/super_ser_rmse_pos.txt};

\addplot[color=red!70!black, mark=square, mark options={scale=0.8}] table {result_figures/data_isac/super_1e3_ser_rmse_pos.txt};

\addplot[color=green!60!black, mark=*, dashdotted, mark options={scale=0.8}] table {result_figures/data_isac/ssl_1e3_ser_rmse_pos.txt};

\end{axis}

\end{tikzpicture}
    \caption{ISAC performance under hardware impairments. Only Pareto optimal points are shown. SL: supervised learning. SSL: semi-supervised learning.}
    \label{fig_isac_results}
\end{figure}

Fig.~\ref{fig_isac_results} shows the \ac{ISAC} results when communication transmission is added, and the joint precoder is computed according to \eqref{eq_isac_precoder_bench}. Only the values of $\rho$ and $\phi$ in \eqref{eq_isac_precoder_bench} that provide Pareto optimal points are shown. 
Results in Fig.~\ref{fig_isac_results} indicate that SL with 85,000 iterations performs similarly to the baseline with perfectly known impairments. In practical scenarios where the amount of labeled data is constrained, Fig.~\ref{fig_isac_results} shows that SL with 1,000 iterations suffers in performance. On the other hand, using SSL initialized from the limited-data SL yields similar performance to fully SL with 85,000 iterations, without any additional labeled data.

\section{Conclusions} \label{sec_conclusions}
In this work, we studied unsupervised and semi-supervised learning in the context of \ac{ISAC}, under inter-antenna spacing impairments. We proposed an unsupervised loss function that accounted for the impairments. The results showed that relying solely on UL with the proposed loss function does not achieve similar performance as SL. 
When fewer labeled data samples are available, limited-data SL degraded in terms of misdetection probability and symbol error rate. However, SSL, using the limited-data SL as starting point, showed to perform similarly to perfect impairment knowledge and unlimited-data SL. 
The current work is limited to single-target sensing, and future work considers UL tailored to multiple targets.

\balance

\bibliographystyle{IEEEtran}
\bibliography{references}

\begin{thebibliography}{10}
\providecommand{\url}[1]{#1}
\csname url@samestyle\endcsname
\providecommand{\newblock}{\relax}
\providecommand{\bibinfo}[2]{#2}
\providecommand{\BIBentrySTDinterwordspacing}{\spaceskip=0pt\relax}
\providecommand{\BIBentryALTinterwordstretchfactor}{4}
\providecommand{\BIBentryALTinterwordspacing}{\spaceskip=\fontdimen2\font plus
\BIBentryALTinterwordstretchfactor\fontdimen3\font minus \fontdimen4\font\relax}
\providecommand{\BIBforeignlanguage}[2]{{%
\expandafter\ifx\csname l@#1\endcsname\relax
\typeout{** WARNING: IEEEtran.bst: No hyphenation pattern has been}%
\typeout{** loaded for the language `#1'. Using the pattern for}%
\typeout{** the default language instead.}%
\else
\language=\csname l@#1\endcsname
\fi
#2}}
\providecommand{\BIBdecl}{\relax}
\BIBdecl

\bibitem{cui2023integrated}
Y.~Cui, F.~Liu, C.~Masouros, J.~Xu, T.~X. Han, and Y.~C. Eldar, ``Integrated sensing and communications: Background and applications,'' in \emph{Integrated Sensing and Communications}.\hskip 1em plus 0.5em minus 0.4em\relax Springer, 2023, pp. 3--21.

\bibitem{wymeersch2021integration}
H.~Wymeersch, D.~Shrestha, C.~M. De~Lima, V.~Yajnanarayana, B.~Richerzhagen, M.~F. Keskin, K.~Schindhelm, A.~Ramirez, A.~Wolfgang, M.~F. De~Guzman \emph{et~al.}, ``Integration of communication and sensing in {6G}: A joint industrial and academic perspective,'' in \emph{Proc. 32nd IEEE Annu. Int. Symp. Personal Indoor Mobile Radio Commun. (PIMRC)}, Helsinki, Finland, 2021, pp. 1--7.

\bibitem{liu2022integrated}
F.~Liu, Y.~Cui, C.~Masouros, J.~Xu, T.~X. Han, Y.~C. Eldar, and S.~Buzzi, ``Integrated sensing and communications: Towards dual-functional wireless networks for {6G} and beyond,'' \emph{IEEE J. Sel. Areas Commun.}, vol.~40, no.~6, pp. 1728--1767, Mar. 2022.

\bibitem{lu2023integrated}
S.~Lu, F.~Liu, Y.~Li, K.~Zhang, H.~Huang, J.~Zou, X.~Li, Y.~Dong, F.~Dong, J.~Zhu \emph{et~al.}, ``Integrated sensing and communications: Recent advances and ten open challenges,'' \emph{arXiv preprint arXiv:2305.00179}, 2023.

\bibitem{liyanaarachchi2021joint}
S.~D. Liyanaarachchi, C.~B. Barneto, T.~Riihonen, M.~Heino, and M.~Valkama, ``Joint multi-user communication and {MIMO} radar through full-duplex hybrid beamforming,'' in \emph{Proc. 1st IEEE Int. Online Symp. Joint Commun. \& Sens. (JC\&S)}, Dresden, Germany, 2021, pp. 1--5.

\bibitem{dokhanchi2021adaptive}
S.~H. Dokhanchi, M.~B. Shankar, M.~Alaee-Kerahroodi, and B.~Ottersten, ``Adaptive waveform design for automotive joint radar-communication systems,'' \emph{IEEE Trans. Veh. Technol.}, vol.~70, no.~5, pp. 4273--4290, Apr. 2021.

\bibitem{OFDM_DFRC_TSP_2021}
M.~F. Keskin, V.~Koivunen, and H.~Wymeersch, ``Limited feedforward waveform design for {OFDM} dual-functional radar-communications,'' \emph{IEEE Trans. Signal Process.}, vol.~69, pp. 2955--2970, Apr. 2021.

\bibitem{chen2021joint}
L.~Chen, F.~Liu, W.~Wang, and C.~Masouros, ``Joint radar-communication transmission: A generalized pareto optimization framework,'' \emph{IEEE Trans. Signal Process.}, vol.~69, pp. 2752--2765, May 2021.

\bibitem{johnston2022mimo}
J.~Johnston, L.~Venturino, E.~Grossi, M.~Lops, and X.~Wang, ``{MIMO} {OFDM} dual-function radar-communication under error rate and beampattern constraints,'' \emph{IEEE J. Select. Areas Commun.}, vol.~40, no.~6, pp. 1951--1964, Mar. 2022.

\bibitem{tan2021integrated}
D.~K.~P. Tan, J.~He, Y.~Li, A.~Bayesteh, Y.~Chen, P.~Zhu, and W.~Tong, ``Integrated sensing and communication in {6G}: Motivations, use cases, requirements, challenges and future directions,'' in \emph{Proc. 1st IEEE Int. Symp. Joint Commun. \& Sens. (JC\&S)}, Dresden, Germany, 2021, pp. 1--6.

\bibitem{chowdhury20206g}
M.~Z. Chowdhury, M.~Shahjalal, S.~Ahmed, and Y.~M. Jang, ``{6G} wireless communication systems: Applications, requirements, technologies, challenges, and research directions,'' \emph{IEEE Open J. Commun. Soc.}, vol.~1, pp. 957--975, Jul. 2020.

\bibitem{jiang2021road}
W.~Jiang, B.~Han, M.~A. Habibi, and H.~D. Schotten, ``The road towards {6G}: A comprehensive survey,'' \emph{IEEE Open J. Commun. Soc.}, vol.~2, pp. 334--366, Feb. 2021.

\bibitem{liu2022learning}
C.~Liu, W.~Yuan, S.~Li, X.~Liu, H.~Li, D.~W.~K. Ng, and Y.~Li, ``Learning-based predictive beamforming for integrated sensing and communication in vehicular networks,'' \emph{IEEE J. Sel. Areas Comm.}, vol.~40, no.~8, pp. 2317--2334, Jun. 2022.

\bibitem{wu2023sensing}
Y.~Wu, F.~Lemic, C.~Han, and Z.~Chen, ``Sensing integrated {DFT}-spread {OFDM} waveform and deep learning-powered receiver design for terahertz integrated sensing and communication systems,'' \emph{IEEE Trans. Commun.}, vol.~71, no.~1, pp. 595--610, Jan. 2023.

\bibitem{liu2023deep}
Y.~Liu, I.~Al-Nahhal, O.~A. Dobre, and F.~Wang, ``Deep-learning channel estimation for {IRS}-assisted integrated sensing and communication system,'' \emph{IEEE Trans. Veh. Technology}, vol.~72, no.~5, pp. 6181--6193, May 2023.

\bibitem{OShea2017}
T.~O'Shea and J.~Hoydis, ``{An Introduction to Deep Learning for the Physical Layer},'' \emph{IEEE Trans. Cogn. Commun. Netw.}, vol.~3, no.~4, pp. 563--575, dec 2017.

\bibitem{mateos2022end}
J.~M. Mateos-Ramos, J.~Song, Y.~Wu, C.~H{\"a}ger, M.~F. Keskin, V.~Yajnanarayana, and H.~Wymeersch, ``End-to-end learning for integrated sensing and communication,'' in \emph{Proc. IEEE Int. Conf. Commun. (ICC)}, Seoul, Korea, Republic of, 2022, pp. 1942--1947.

\bibitem{muth2023autoencoder}
C.~Muth and L.~Schmalen, ``Autoencoder-based joint communication and sensing of multiple targets,'' in \emph{Proc. 26th VDE Int. ITG Workshop Smart Antennas and Conf. Syst., Commun., Coding}, Braunschweig, Germany, 2023, pp. 1--6.

\bibitem{Shlezinger2023}
N.~Shlezinger, J.~Whang, Y.~C. Eldar, and A.~G. Dimakis, ``Model-based deep learning,'' \emph{Proceedings of the IEEE}, vol. 111, no.~5, pp. 465--499, 2023.

\bibitem{Xiuhong21}
X.~Wei, C.~Hu, and L.~Dai, ``Deep learning for beamspace channel estimation in millimeter-wave massive {MIMO} systems,'' \emph{IEEE Trans. Commun.}, vol.~69, no.~1, pp. 182--193, Jan. 2021.

\bibitem{Hengtao18}
H.~He, C.-K. Wen, S.~Jin, and G.~Y. Li, ``Deep learning-based channel estimation for beamspace {mmWave} massive {MIMO} systems,'' \emph{IEEE Wireless Commun. Lett.}, vol.~7, no.~5, pp. 852--855, Oct. 2018.

\bibitem{chatelier2023modelbased}
B.~Chatelier, L.~L. Magoarou, V.~Corlay, and M.~Crussière, ``Model-based learning for location-to-channel mapping,'' \emph{arXiv preprint arXiv:2308.14370}, 2023.

\bibitem{xiao2020deepfpc}
P.~Xiao, B.~Liao, and N.~Deligiannis, ``{DeepFPC}: A deep unfolded network for sparse signal recovery from 1-bit measurements with application to {DOA} estimation,'' \emph{Signal Process.}, vol. 176, p. 107699, Nov. 2020.

\bibitem{wu2022doa}
L.~Wu, Z.~Liu, and J.~Liao, ``{DOA} estimation using an unfolded deep network in the presence of array imperfections,'' in \emph{Proc. 7th IEEE International Conf. on Signal and Image Processing (ICSIP)}, Suzhou, China, 2022, pp. 182--187.

\bibitem{mateos2023model}
J.~M. Mateos-Ramos, C.~H{\"a}ger, M.~F. Keskin, L.~L. Magoarou, and H.~Wymeersch, ``Model-based end-to-end learning for multi-target integrated sensing and communication,'' \emph{arXiv preprint arXiv:2307.04111}, 2023.

\bibitem{chapelle2006ssl}
O.~Chapelle, B.~Scholkopf, and A.~Zien, \emph{Semi-Supervised Learning}.\hskip 1em plus 0.5em minus 0.4em\relax MIT press, 2006.

\bibitem{yassine2022}
T.~Yassine and L.~Le~Magoarou, ``{mpNet}: Variable depth unfolded neural network for massive {MIMO} channel estimation,'' \emph{IEEE Trans. Wireless Commun.}, vol.~21, no.~7, pp. 5703--5714, Jan. 2022.

\bibitem{Chatelier2022}
B.~Chatelier, L.~Le~Magoarou, and G.~Redieteab, ``Efficient deep unfolding for {SISO-OFDM} channel estimation,'' in \emph{Proc. IEEE Int. Conf. Commun. (ICC)}, Rome, Italy, 2023.

\bibitem{OFDM_radar_TVT_2020}
J.~B. Sanson, P.~M. Tomé, D.~Castanheira, A.~Gameiro, and P.~P. Monteiro, ``High-resolution delay-doppler estimation using received communication signals for {OFDM} radar-communication system,'' \emph{IEEE Trans. Veh. Technology}, vol.~69, no.~11, pp. 13\,112--13\,123, Sep. 2020.

\bibitem{MIMO_OFDM_ICI_JSTSP_2021}
M.~F. Keskin, H.~Wymeersch, and V.~Koivunen, ``{MIMO-OFDM} joint radar-communications: Is {ICI} friend or foe?'' \emph{IEEE J. of Select. Topics Signal Process.}, vol.~15, no.~6, pp. 1393--1408, Sep. 2021.

\bibitem{5G_NR_JRC_analysis_JSAC_2022}
L.~Pucci, E.~Paolini, and A.~Giorgetti, ``System-level analysis of joint sensing and communication based on {5G} new radio,'' \emph{IEEE J. Select. Areas Commun.}, vol.~40, no.~7, pp. 2043--2055, Mar. 2022.

\bibitem{overview_SP_JCS_JSTSP_2021}
J.~A. Zhang, F.~Liu, C.~Masouros, R.~W. Heath, Z.~Feng, L.~Zheng, and A.~Petropulu, ``An overview of signal processing techniques for joint communication and radar sensing,'' \emph{IEEE J. Select. Topics Signal Process.}, vol.~15, no.~6, pp. 1295--1315, Sep. 2021.

\bibitem{zhang2018multibeam}
J.~A. Zhang, X.~Huang, Y.~J. Guo, J.~Yuan, and R.~W. Heath, ``Multibeam for joint communication and radar sensing using steerable analog antenna arrays,'' \emph{IEEE Trans. Veh. Technol.}, vol.~68, no.~1, pp. 671--685, Nov. 2018.

\bibitem{schenk2008rf}
T.~Schenk, \emph{RF imperfections in high-rate wireless systems: impact and digital compensation}.\hskip 1em plus 0.5em minus 0.4em\relax Springer Science \& Business Media, 2008.

\bibitem{chen2023modeling}
H.~Chen, M.~F. Keskin, S.~R. Aghdam, H.~Kim, S.~Lindberg, A.~Wolfgang, T.~E. Abrudan, T.~Eriksson, and H.~Wymeersch, ``Modeling and analysis of {6G} joint localization and communication under hardware impairments,'' \emph{arXiv preprint arXiv:2301.01042}, 2023.

\bibitem{precoding_mmWave_JSTSP_2014}
A.~Alkhateeb, O.~El~Ayach, G.~Leus, and R.~W. Heath, ``Channel estimation and hybrid precoding for millimeter wave cellular systems,'' \emph{IEEE J. Sel. Topics Signal Process.}, vol.~8, no.~5, pp. 831--846, Jul. 2014.

\bibitem{analogBeamformerDesign_TSP_2017}
J.~Tranter, N.~D. Sidiropoulos, X.~Fu, and A.~Swami, ``Fast unit-modulus least squares with applications in beamforming,'' \emph{IEEE Trans. Signal Process.}, vol.~65, no.~11, pp. 2875--2887, Feb. 2017.

\bibitem{OFDM_Radar_Corr_TAES_2020}
S.~Mercier, S.~Bidon, D.~Roque, and C.~Enderli, ``Comparison of correlation-based {OFDM} radar receivers,'' \emph{IEEE Trans. Aerospace Electron. Syst.}, vol.~56, no.~6, pp. 4796--4813, Jun. 2020.

\bibitem{reciprocalFilter_OFDM_2023}
J.~T. Rodriguez, F.~Colone, and P.~Lombardo, ``Supervised reciprocal filter for {OFDM} radar signal processing,'' \emph{IEEE Trans. Aerospace Electron. Syst.}, pp. 1--22, Jan. 2023.

\bibitem{MAP_Detector_TSP_2021}
S.~Guruacharya, B.~K. Chalise, and B.~Himed, ``{MAP} ratio test detector for radar system,'' \emph{IEEE Trans. Signal Process.}, vol.~69, pp. 573--588, Dec. 2021.

\bibitem{richards2005fundamentals}
M.~A. Richards, \emph{Fundamentals of Radar Signal Processing}.\hskip 1em plus 0.5em minus 0.4em\relax Tata McGraw-Hill Education, 2005.

\bibitem{mateos2022model}
J.~M. Mateos-Ramos, C.~H{\"a}ger, M.~F. Keskin, L.~Le~Magoarou, and H.~Wymeersch, ``Model-driven end-to-end learning for integrated sensing and communication,'' in \emph{Proc. IEEE Int. Conf. Commun. (ICC)}, Rome, Italy, 2023.

\bibitem{kingma2015adam}
D.~P. Kingma and J.~Ba, ``Adam: A method for stochastic optimization,'' in \emph{Proc. 3rd Int. Conf. Learn. Representations (ICLR)}, San Diego, CA, USA, 2015.

\end{thebibliography}

\end{document}